\documentclass[%
 reprint,
 amsmath,amssymb,
 aps,
]{revtex4-1}

\usepackage{units}
\usepackage{siunitx}
\usepackage{graphicx}
\usepackage{dcolumn}
\usepackage{bm}

\begin{document}

\preprint{APS/123-QED}

\title{Ultrafast Energy Transfer with Competing Channels: \\
Non-equilibrium F\"{o}rster and Modified Redfield Theories}%

\author{Joachim Seibt}
\email{seibt@karlov.mff.cuni.cz}
\author{Tom\'{a}\v{s} Man\v{c}al}%
\affiliation{%
 Faculty of Mathematics and Physics, Ke Karlovu 5, 121 16 Prague 2, Czech Republic
}%

\date{\today}

\begin{abstract}
We derive equations of motion for the reduced density matrix of a molecular system which undergoes energy transfer dynamics competing with fast internal conversion channels. Environmental degrees of freedom of such a system have no time to relax to quasi-equilibrium in the electronic excited state of the donor molecule, and thus the conditions of validity of F\"{o}rster and Modified Redfield theories in their standard formulations do not apply. We derive non-equilibrium versions of the two well-known rate theories and apply them to the case of carotenoid-chlorophyll energy transfer. Although our reduced density matrix approach does not account for the formation of vibronic excitons, it still confirms the important role of the donor ground-state vibrational states in establishing the resonance energy transfer conditions. We show that it is essential to work with a theory valid in strong system-bath interaction regime to obtain correct dependence of the rates on donor-acceptor energy gap.    
\end{abstract}

\pacs{Valid PACS appear here}
\maketitle

\section{\label{sec:intro}Introduction}

Excitation energy transfer (EET) processes represent an important
subclass of transport phenomena in open quantum systems \cite{May2000}.
Studies of EET processes in condensed matter, artificial nano-materials
and biological systems inspire applied renewable energy research,
as well as provide fundamental insights into important natural processes.
The latter is especially true for biological research. In the field
of photosynthesis, soon after it was recognized in 1990s that 
closely packed aggregates of (bacterio)chlorophylls are responsible for
light energy harvesting and initial energy transport, EET processes became
the prime subject of theoretical studies. A consistent picture of
EET dynamics in photosynthetic molecular aggregates has emerged
towards the end of past century \cite{VanAmerongen2000,May2000}.
The field has drawn heavily from the previous advances in experimental
laser science \cite{Fleming1986,Demtroder2008}, the theory of time
resolved spectroscopy \cite{Mukamel1995Book}, the theory of transfer
processes in molecular crystals \cite{Silinsh1994} and the whole
development in the field of dissipative dynamics \cite{Weiss2008,Fain2000}.

Master equations for the reduced density matrix (RDM), mostly in Markovian,
but also in non-Markovian formulations \cite{Cerrillo2014}, often in conjugation
with response function theory of non-linear ultrafast spectroscopy \cite{Mukamel1995Book},
form the basis of most of the successful theories of EET processes in photosynthesis 
\cite{Chenu2015}. Depending on the type
of the molecular system, various types of master equations describe the EET
processes successfully. In general, there are two limits in which
master equations provide correct dynamics. Both these limits are defined
by the region of validity of certain perturbation theories. If, for
instance, the resonance coupling between two molecular systems can
be considered a small parameter, one arrives at a set of RDM
equations containing rates similar to the well-known F\"{o}rster
resonance energy transfer rates \cite{Foerster1946,Foerster1948}.
If, on the other hand, the system--bath coupling can be considered
a small parameter, one arrives at master equations related to those
proposed originally by A. G. Redfield in the framework of nuclear
magnetic resonance \cite{Redfield1965}.

The original result of F\"{o}rster can be conveniently written in
terms of experimentally accessible quantities (see e. g. \cite{VanAmerongen2000})
and it has gained substantial popularity due to its intuitive character.
In molecular complexes, simple predictions of the F\"{o}rster theory
often fail \cite{Mukai1999,Scholes2000,Zigmantas2006}, but generalizations
of the original ideas to interactions between the whole molecular complexes is
straightforward \cite{Mukai1999,Scholes2001,Jang2004}. Similarly,
Redfield equations adapted for molecular aggregates proved to be an
extremely versatile tool \cite{VanAmerongen2000,May2000,Valkunas2013}.
Frenkel exciton model in conjugation with the Redfield equations, often
in combination with the F\"{o}rster theory, is behind a substantial part of the
qualitative insight we have into the inner workings of photosynthesis (see e.g. \cite{Novoderezhkin2004,Novoderezhkin2011,Zigmantas2006,Cho2005,Adolphs2006}).
The two general theories described above are often stretched beyond their
formal region of validity, even though they rarely cease to deliver
meaningful physical insights.

Despite current attention to fine effects of underdamped intramolecular vibrational modes 
\cite{Christensson2012,Tiwari2013,Chin2013,Womick2011,Butkus2014a} and to details of spectral density shapes \cite{Kreisbeck2012}, and despite recent efforts
to introduce a paradigm shift in understanding of the origin of the photosynthetic EET efficiency \cite{Ball2011,Scholes2010,Wolynes2009},
the overall picture of this process, as drawn by the master equations of the F\"{o}rster and
Redfield types, remains valid \cite{Strumpfer2012} (see e. g. \cite{Wilkins2015}
for quantitative results). Recent theoretical advances enabling exact numerical solutions of some types of energy transfer problems e.g. by Hierarchical equations of motion (HEOM) \cite{Ishizaki2009,Ishizaki2009a,Ishizaki2009b} or by the Time evolving density with orthogonal polynomial algorithm (TEDOPA) approach \cite{Oviedo-Casado2016,Rosenbach2016}, also do not change the picture qualitatively, but rather improve quantitative aspects of our understanding.

The approximate F\"{o}rster/Redfield
theories will continue to play an important role in our understanding
of EET phenomena for years to come. A testimonial to this are recent
works attempting to improve their accuracy in various regimes of approximation \cite{Jang2002,Jang2008,Jang2011,Banchi2013,Kimura2016c,Chang2013,Sun2016a}. 
Our effort, presented in this work, can be understood as an integral part of this trend.

We are motivated by our recent study of ultrafast energy transfer
between the bright $S_{2}$ state of a carotenoid molecule and the
second lowest excited state, the so-called $Q_{x}$ state, of (bacterio)chlorophyll
((B)Chl) and related molecules \cite{Perlik2015}. By means of an explicit
non-perturbative treatment of carotenoid vibrational modes, it was
possible to show in Ref. \cite{Perlik2015} that multi-vibrational-quantum
transitions on the carotenoid are responsible for the measured sub
100 fs energy transfer time from carotenoid to chlorophyll. This
transition is so fast that it outcompetes even the internal relaxation from
$S_{2}\rightarrow S_{1}$ in the carotenoid. 

Any such explicit inclusion
of vibrational degrees of freedom (DOF) into the Hamiltonian becomes
quickly numerically too expensive when the size of the molecular system
grows. It would therefore be of interest to see if methods treating
nuclear DOF as a bath can reproduce the above mentioned ultrafast
EET rates. The multi-quantum character of the transitions suggests
that any theory of energy transfer in this system has to go beyond
second order in system-bath interaction. This excludes ordinary Redfield
theory from consideration, and suggests F\"{o}rster and Modified Redfield
theories as the possible candidates, due to their (partial in case
of Modified Redfield) non-perturbative character with respect to system--bath
coupling. In their standard formulations (see e.g. \cite{Valkunas2013}
Chap. 11.5 and 11.6) it is assumed that the state of the bath corresponding
to the initial condition of the EET is adapted to the excited state
potential energy surfaces (PES). This assumption works reasonably well for the
F\"{o}rster theory and slow EET rates, where the initial equilibration
of the bath in the excited state occurs much faster than the subsequent/competing
EET. In such a formulation the equations are, strictly speaking, valid
on time-scales on which short time processes, such as bath reorganization
or dephasing of coherences, are already over. In case of Modified
Redfield theory we find the same limitation. Modified Redfield theory
was, moreover, derived only for populations of delocalized excitonic
levels \cite{Zhang1998,Yang2002}. While the reason for not treating
coherence elements of the density matrix is technical (the derivation
within the projection operator technique is only possible with an
operator projecting on the diagonal elements of the density matrix
only), in the time interval in which the theory is valid, the coherence
elements of the RDM should be zero anyway. Standard formulations of
the F\"{o}rster and Modified Redfield theories are therefore not suitable
for the problem of $S_{2}\rightarrow Q_{x}$ transition, because the
bath DOF of the carotenoid, when the system is excited to $S_{2}$,
have no time to equilibriate. Not only the transfer from $S_{2}$ to
$Q_{x}$ occurs fast, also the depopulation of $S_{2}$ due to other
competing channels is ultrafast. It is therefore much more reasonable
to assume that the bath is close to the state in which it was right
after the excitation (experiment is performed with ultrashort pulses).
This excited bath state corresponds, in Condon approximation, to the bath equilibrium
established prior to excitation on the electronic ground-state. On the excited state PES,
such initial state of the bath represents a highly non-equilibrium
state, and it is contrasted here with the usual electronic excited state
equilibrium assumed in the standard formulations of the two relaxation
theories. Our task is therefore to derive the non-equilibrium equivalents
of the F\"{o}rster and Modified Redfield theories.

The paper is organized as follows: First the theoretical background is reviewed in Sec.~\ref{sec:theoretical_background}.
It starts with a derivation of a quantum master equation in Sec.~\ref{sec:deriv_QME} using the projection operator formalism under the assumption that the bath is not in thermal equilibrium initially. Then the general expression is specified for the cases of F\"{o}rster and Modified Redfield population transfer in Sec.~\ref{sec:Foerster_transfer} and \ref{sec:modified_Redfield_transfer}, respectively.
The results of model calculations
, in particular 
a comparison of the time-dependencies of F\"{o}rster and Modified Redfield rates obtained from standard and non-equilibrium approaches,
extraction of transfer rates from the population dynamics of a system with additional relaxation channels, and effects of finite excitaion pulse width,
are discussed in Sec.~\ref{sec:results_discussion}.
Special attention is paid to the donor--acceptor energy gap dependence of the transfer rates which is investigated in Sec.~\ref{sec:frequency_dependence_transfer_rates}.
An overview of the main aspects of this article is given in Conclusions, Sec.~\ref{sec:Conclusions}.

\section{Theoretical background} \label{sec:theoretical_background}

\subsection{Derivation of quantum master equation using projection operator formalism} \label{sec:deriv_QME}

Let us consider a dimer donor--acceptor system which undergoes population transfer subsequent to electronic excitation from ground- to singly excited state of the donor. We assume that the initial state of the donor can be factorized into an electronic part, represented by the electronic ground-state of the donor,
and an equilibrium state of the bath. The bath includes both the intramolecular nuclear DOF of the donor molecule and the degrees of freedom (DOF) of the environment of the donor. The factorization assumption is reasonable as we assume that the energy gap is optical, and no thermal electronic excitation can therefore exist. The effect of electronic excitation on the bath DOF coincides with a displacement of the bath oscillators. 
The total Hamiltonian of the dimer (including its environmental DOF) can be decomposed into a reference Hamiltonian $H^{0}$ and a perturbation Hamiltonian $H'$. Different choices of the reference and perturbation Hamiltonians enable us to apply perturbation theory with validity in different ranges of system parameters.

The Liouville equation for time evolution of the density matrix $\rho$ is formulated in the interaction picture with respect to $H'$ with Liouville operator ${\cal L}_I(t)$ including the time-dependence of the perturbation Hamiltonian under the influence of the reference Hamiltonian $H'$:
\begin{equation}
\dot{\rho}(t)=-i {\cal L}_I(t) \rho(t)=-i \left[ H'(t), \rho(t) \right],
\end{equation}
where $H'(t)=\exp(-iH^0t/\hbar)H'\exp(iH^0t/\hbar)$.
To facilitate a systematic treatment of the combined dynamics of system and bath, the bath degrees of freedom can be traced out by applying a projection operator \cite{Yang2002}. 
This projection operator includes the projection on the eigenstates $|a\rangle$ of the system, 
the bath density matrix in thermal equilibrium $\rho_{b,eq,g}$ of the electronic ground-state and the trace over the bath $Tr_q$ in terms of
\begin{equation} \label{eq:projection_operator}
{\cal P} A=\sum_a \rho_{b,eq,g} | a \rangle \langle a | Tr_q \{ A_{aa} \}. 
\end{equation}
Furthermore, by defining a complementary projection operator ${\cal Q}=1-{\cal P}$ and by inserting the identity ${\cal P}+{\cal Q}$ after ${\cal L}_I$ \cite{May2000},
we obtain coupled differential equations
\begin{equation} \label{eq:Liouville_equation_projectors_P_Q}
\begin{split}
{\cal P} \dot{\rho}(t)&=-i {\cal P} {\cal L}_I(t) ({\cal P}+{\cal Q}) \rho(t), \\
{\cal Q} \dot{\rho}(t)&=-i {\cal Q} {\cal L}_I(t) ({\cal P}+{\cal Q}) \rho(t).
\end{split}
\end{equation} 
Inserting the solution of the second component of Eq.~\ref{eq:Liouville_equation_projectors_P_Q} into the first component yields the Nakajima-Zwanzig identity (see e.g. Ref. \cite{Valkunas2013})
\begin{equation} \label{eq:Nakajima_Zwanzig_identity}
{\cal P} \dot{\rho}(t)=-i {\cal I}_{NZ}(t)-i {\cal L}_{NZ}(t) {\cal P} \rho(t)-{\cal K}_{NZ}(t,{\cal P} \rho),
\end{equation} 
where in contrast to the terms ${\cal I}_{NZ}(t)$ and ${\cal L}_{NZ}(t)$ the term ${\cal K}_{NZ}(t,{\cal P} \rho)$ includes a convolution of the time evolution of the density matrix 
with a memory kernel.
However, Eq.~(\ref{eq:Nakajima_Zwanzig_identity}) can be recast into a convolutionless form 
\begin{equation} \label{eq:Nakajima_Zwanzig_identity_convolution_free}
{\cal P} \dot{\rho}(t)=-i {\cal I}_{CL}(t)-{\cal K}_{CL}(t) {\cal P} \rho(t),
\end{equation} 
which after taking the trace corresponds to
\begin{equation} \label{eq:QME_solution}
\begin{split}
&\sum_a | a \rangle \langle a | Tr_q \{ \dot{\rho}_{aa}(t) \}= \\ 
&-i \sum_a | a \rangle \langle a | Tr_q \{ ({\cal L}_I(t) \rho_{b,eq,g})_{aa} \} 
\sigma_{I}(0) \\
&-\int_0^t d\tau \sum_a | a \rangle \langle a | Tr_q \{ ({\cal L}_I(t) {\cal Q} {\cal L}_I(\tau) \rho_{b,eq,g})_{aa} \} \sigma_{I}(t).
\end{split}
\end{equation}
By identifying the remaining terms as
\begin{equation} \label{eq:QME_R-term}
\begin{split}
&{\cal R}(t)={\cal K}_{CL}(t) \\
&=\int_0^t d\tau \sum_a | a \rangle \langle a | Tr_q \{ ({\cal L}_I(t) {\cal Q} {\cal L}_I(\tau) \rho_{b,eq,g})_{aa} \},
\end{split}
\end{equation}
and
\begin{equation} \label{eq:QME_I-term_first_order}
{\cal I}(t)=-i {\cal I}_{CL}(t)=-i \sum_a | a \rangle \langle a | Tr_q \{ ({\cal L}_I(t) \rho_{b,eq,g})_{aa} \} \sigma_{I}(0),
\end{equation}
the quantum master equation (QME) for the system density matrix $\sigma_I=\sum_a | a \rangle \langle a | Tr_q \{ \rho_{aa} \}$ becomes \cite{Jang2008}
\begin{equation} \label{eq:QME_system}
\dot{\sigma}_I(t)=-{\cal R}(t)\sigma_I(t)+{\cal I}(t).
\end{equation}
Note that $\sigma_I(t)$ remains diagonal during the time evolution because all operators in Eq.~\ref{eq:QME_system} are diagonal by definition.
Under the assumption that the perturbation Hamiltonian $H'$ entering in ${\cal L}_I$ is off-diagonal in the basis of the electronic eigenstates, it is possible to formulate ${\cal R}(t)$ as a product of two components. One of them is a population transfer superoperator \cite{Seibt2014} acting only on the system density matrix, and the other one accounts for the influence of the bath. 
The inhomogeneous term ${\cal I}(t)$ includes the commutator of the electronic component of $H'$ with $\sigma_I(0)$.

\subsection{Calculation of F\"{o}rster transfer rates} \label{sec:Foerster_transfer}

For the case of F\"{o}rster transfer between a donor and an acceptor, where the resonance Coulomb couplings $J_{nm}$ between the molecular transitions on molecules $n$ and $m$ are small compared to the system-bath coupling, the matrix elements of the perturbation Hamiltonian read as
\begin{equation}
H'_{mn}=J_{m n} |m \rangle \langle n |.
\end{equation}
Besides the standard description F\"{o}rster theory, a non-equilibrium generalization has been derived previously \cite{Jang2002}, however without applying the cumulant expansion to obtain a compact line shape function based formulation.
We follow the derivation steps given in \cite{Yang2002} under the modified assumption that the bath in the excited electronic state is not equilibrated initially. It rather corresponds to the equilibrium bath related to the electronic ground state.
To formulate rate expressions for further evaluation via cumulant expansion, Eq.~(\ref{eq:QME_solution}) is expressed in terms of tensor elements accounting for population transfer between an initially populated donor molecule (index $m$) and an acceptor (index $n$) as
\begin{equation} \label{eq:Foerster_rate_homogeneous_term_matrix_elements_general_expression}
\begin{split}
&{\cal R}_{nn mm}(t)=2 \Re \int^{t}_{0} {\rm d}t Tr_q \{ \exp(i H_{m}^{0} t) H'_{m n} \\
&\times \exp(-i H_{n}^{0} t) \exp(i H_{n}^{0} t') H'_{n m} \exp(-i H_{m}^{0} t') \rho_{b,eq,g} \}.
\end{split}
\end{equation}
Different from Eq.~(\ref{eq:QME_R-term}), in Eq.~(\ref{eq:Foerster_rate_homogeneous_term_matrix_elements_general_expression}) we take twice the real part of the expression, as the complex conjugate tensor element yields complex conjugated contribution.
Furthermore, in deriving Eq.~(\ref{eq:Foerster_rate_homogeneous_term_matrix_elements_general_expression}) we assume that only the unity operator entering in ${\cal Q}$ remains, as the term containing ${\cal P}$ becomes zero.
Because of the off-diagonal form of the perturbation Hamiltonian and the initially diagonal form of the density matrix, the selection of diagonal elements by the projection operator given in Eq.~(\ref{eq:projection_operator}) makes the respective expression vanish.
The Hamiltonian operators $H_k$ of the donor ($k=m \in \{ 1,2 \}$) and the acceptor ($k=n \in \{ 1,2 \}, n \neq m$) contain electronic excitation energies $e^0_k$, reorganization energies $l_k$, bath phonon energies $e^{ph}_k$ and energy gap coordinates $u_k$ associated with system-bath coupling \cite{Yang2002}. 
The bath component of ${\cal R}_{nn mm}(t)$ can be formulated in terms of time-ordered exponentials containing integrals over energy gap coordinates in the interaction picture (see e.g. \cite{Mukamel1995Book})
\begin{equation} \label{eq:time_evolution_energy_gap_coordinates}
  u_k(\tau)=\exp(i e^{ph}_k \tau) u_k \exp(-i e^{ph}_k \tau).
\end{equation}
In the framework of the second-order cumulant expansion, the line shape functions can be identified as
\begin{equation} \label{eq:line_shape_function_correlation_function}
  g_{k}(\tau)=\int_{0}^{\tau} d\tau' \int_{0}^{\tau'} d\tau'' Tr_q \{ u_k(\tau'') u_k(0) \}
\end{equation}
under the assumption that the bath fluctuations associated with the singly excited states of donor and acceptor can be considered as uncorrelated, so that the cumulant expansion 
only yields line shape functions related to either the donor or the acceptor.
Eq.~(\ref{eq:QME_R-term}) then becomes
\begin{equation} \label{eq:Foerster_rate_homogeneous_term_line_shape_functions}
\begin{split}
&{\cal R}_{nn mm,bath,noneq}(t)= 2 |J_{mn}|^2 \\
&\Re \bigg( \int_0^t dt' \exp(i (e^0_m-e^0_n) (t-t')) \exp(i (l_m-l_n) (t-t')) \\
&\exp \left( -g_{n}(t-t')-g_{m}(t-t') \right. \\
&\left. +2i\Im(g_{m}(t))-2i\Im(g_{m}(t')) \right) \bigg)
\end{split}
\end{equation}
For comparison, the analogous expression in the case of the standard F\"{o}rster approach reads \cite{Yang2002}
\begin{equation} \label{eq:standard_Foerster_rate_homogeneous_term_line_shape_functions}
\begin{split}
&{\cal R}_{nn mm,bath,std}(t)= 2 |J_{mn}|^2 \\
&\Re \bigg( \int_0^t dt' \exp(i (e^0_m-e^0_n) (t-t')) \exp(i (-l_m-l_n) (t-t')) \\
&\exp \left( -g_{n}(t-t')-g_{m}(t-t') \right) \bigg)
\end{split}
\end{equation}
The F\"{o}rster transfer rate can also be expressed in terms of an integral over the product of absorption and complex conjugate emission component, which in the case of the standard description correspond to
\begin{equation}
A_n(t')=\exp(-i (e^0_n+l_n) t') \exp \left(-g_{n}(t')\right)
\end{equation}
and
\begin{equation}
F_m(t')=\exp(-i (e^0_m-l_m) t') \exp \left(-g^{*}_{m}(t')\right)
\end{equation}
respectively.
The rate expression then becomes
\begin{equation} \label{eq:standard_Foerster_rate_product_absorption_emission}
{\cal R}_{nn mm,bath,std}(t)= 2 |J_{mn}|^2 \Re \left( \int_0^t dt' F^{*}_m(t') A_n(t') \right).
\end{equation}
Note that in the limit of large time argument $t-t'$, the integrand functions from Eqs.~(\ref{eq:Foerster_rate_homogeneous_term_line_shape_functions}) and (\ref{eq:standard_Foerster_rate_homogeneous_term_line_shape_functions}) become equivalent, because
$\lim_{\tau \to \infty}\dot{g}_k(\tau)=-l_k$ \cite{Zhang1998}. However, this limit is not applicable for small values of $t-t'$,
so that the different approaches lead to different results for $t'$ approaching the upper integration border. 

An analogous approach as in the derivation of the homogeneous term given in Eq.~(\ref{eq:Foerster_rate_homogeneous_term_line_shape_functions}) for the formulation of the inhomogeneous term from Eq.~(\ref{eq:QME_I-term_first_order}) leads to
\begin{equation} \label{eq:Foerster_rate_homogeneous_term_second_order}
\begin{split}
&{\cal I}_{nn mm,bath}(t)= 2 J_{mn} \Im \left( \exp(-i (e^0_m-e^0_n) t) \right. \\
&\left. \exp(-i (l_m-l_n) t) \exp(-g^{*}_{n}(t)-g_{m}(t)) \right).
\end{split}
\end{equation}
The evolution of the system can be treated separately by introducing a superoperator for population transfer from state $m$ to state $n$, which can be expressed in terms of matrix elements of an operator $\Theta$ with
\begin{equation} \label{eq:definition_Theta_Foerster}
\Theta_{mn}=|m \rangle \langle n|.
\end{equation}
This off-diagonal operator accounts for the influence of the electronic component of the perturbation Hamiltonian. The matrix elements of the relaxation superoperator are \cite{Seibt2014}
\begin{equation}
{\cal K}_{nnmm} \; \bullet = [\Theta_{mn},\Theta_{nm} \; \bullet - \bullet \; \Theta_{mn}].
\end{equation}
These matrix elements enter in ${\cal R}_{nn mm}(t)={\cal R}_{nn mm,bath}(t) {\cal K}_{nnmm}$. Note that the selection of diagonal elements in the system eigenbasis according to the definition of the projection operator from Eq.~(\ref{eq:projection_operator}) implicitly enters in the relaxation superoperator.
For the inhomogeneous term, one obtains ${\cal I}_{nn mm}(t)=\sum_a | a \rangle \langle a | \left( {\cal I}_{nn mm,bath}(t) [\Theta_{nm}, \sigma_I(0)] \right)_{aa}$.
Selecting diagonal elements from the off-diagonal commutator expressions makes ${\cal I}_{nn mm}(t)$ vanish. The rate equation can then be formulated as
\begin{equation} \label{eq:rate_equation}
  \dot{\sigma}_{I,nn}(t)=-\sum_m {\cal R}_{nnmm}(t)\sigma_{I,mm}(t).
\end{equation}

\subsection{Calculation of Modified Redfield transfer rates} \label{sec:modified_Redfield_transfer}

In the case that the system-bath coupling is sufficiently small, it can be treated in a perturbative way, while the resonance coupling enters in the description of the system via a transformation to the so-called exciton basis. 
The exciton eigenstates $|k\rangle$ are expressed in terms of linear combinations of the localized singly excited states $|n\rangle$ as
\begin{equation} \label{eq:transformation_exciton_basis}
| k \rangle = \sum_n j_{kn} | n \rangle.
\end{equation}
Reference Hamiltonian contains exciton eigenenergies $E_k^0$, phonon energies $e_n^{ph}$, reorganization energies $l_n$ and energy gap coordinates $u_n$, where $l_n$ and $u_n$ are weighted by products of transformation coefficients $a_{k_1 k_2}(n)=j_{k_1 n} j_{k_2 n}, \{ k_1, k_2 \} \in \{ k, k' \}$ with equal indices. The reference Hamiltonian reads as
\begin{equation} \label{eq:reference_Hamiltonian}
H_k^0=\left[ E_k^0 + \sum_n (a_{kk}(n) l_n + e_n^{ph} + a_{kk}(n) u_n) \right] \left| k \rangle \langle k \right|.
\end{equation}
In contrast, the perturbation Hamiltonian
\begin{equation} \label{eq:perturbation_Hamiltonian}
H'_{k k'}=H_{k k'}^{el-ph}=(1-\delta_{k k'}) \left[ \sum_n a_{kk'}(n) u_n \right] \left| k \rangle \langle k' \right|
\end{equation}
contains products of transformation coefficients with different indices.
The rate can be expressed in terms of line shape functions $g_{n,k_1 k_2 k_3 k_4}(\tau)=a_{k_1 k_2}(n) a_{k_3 k_4}(n) g_{n}(\tau)$ and reorganization energies $l_{n,k_1 k_2 k_3 k_4}=a_{k_1 k_2}(n) a_{k_3 k_4}(n) l_{n}$ with the shorthand notations $g_{k_1 k_2 k_3 k_4}(\tau)=\sum_n g_{n,k_1 k_2 k_3 k_4}(\tau)$ and $l_{k_1 k_2 k_3 k_4}=\sum_n l_{n,k_1 k_2 k_3 k_4}$. Details of the derivation are given in the Supporting Information.
As in the case of the F\"{o}rster description, also in the integrand of the Modified Redfield rate expression we can identify an absorption component $A_{k}$ and an emission component $F_{k'}$, the latter taken as complex conjugate. However, there is also an additional component $N_{k k'}$ consisting of line shape function derivatives, so that the rate expression reads as
\begin{equation} \label{eq:modified_Redfield_rate_R}
\begin{split}
&{\cal R}_{kk k'k',bath,noneq}(t)= 2 \Re \left( \phantom{\int} \right. \\
&\left. \int_0^t dt' F^{*}_{k'}(t,t') A_{k}(t,t') N_{k k'}(t,t') \right),
\end{split}
\end{equation}
with
\begin{equation} \label{eq:fluorescence_modified_Redfield_noneq}
\begin{split}
&F_{k'}(t,t')= \exp(-i E_{k'}^0 (t-t')) \\
&\exp(-i l_{k'k'k'k'} (t-t')) \exp \left( -g^{*}_{k'k'k'k'}(t-t') \right. \\
&\left. -2i\Im(g_{k'k'k'k'}(t))+2i\Im(g_{k'k'k'k'}(t')) \right),
\end{split}
\end{equation}
\begin{equation} \label{eq:absorption_modified_Redfield_noneq}
\begin{split}
&A_{k}(t,t')= \exp(-i E_{k}^0 (t-t')) \\
&\exp(-i l_{kkkk} (t-t')) \exp \left( -g_{kkkk}(t-t') \right),
\end{split}
\end{equation}
and
\begin{equation} \label{eq:Nterm_modified_Redfield_rate_noneq}
\begin{split}
&N_{k k'}(t,t')= \exp \left( 2 g_{kkk'k'}(t-t') \right. \\
&\left. -2i \Im(g_{kkk'k'}(t))+2i\Im(g_{kkk'k'}(t')) \right) \\
&\times \{ \ddot{g}_{k'kkk'}(t-t') \\
&-[\dot{g}_{k'kk'k'}(t-t')-\dot{g}_{k'kkk}(t-t') \\
&-2i\Im(\dot{g}_{k'kk'k'}(t))] \times [\dot{g}_{k'k'kk'}(t-t') \\
&-\dot{g}_{kkkk'}(t-t')-2i \Im(\dot{g}_{k'k'kk'}(t'))] \}.
\end{split}
\end{equation}
The standard Modified Redfield rate can be expressed as
\begin{equation} \label{eq:standard_modified_Redfield_rate_R}
\begin{split}
&{\cal R}_{kk k'k',bath,std}(t)= 2 \Re \left( \phantom{\int} \right. \\
&\left. \int_0^t dt' \tilde{F}^{*}_{k'}(t,t') \tilde{A}_{k}(t,t') \tilde{N}_{k k'}(t,t') \right),
\end{split}
\end{equation}
with
\begin{equation} \label{eq:fluorescence_modified_Redfield_std}
\begin{split}
&\tilde{F}_{k'}(t,t')= \exp(-i E_{k'}^0 (t-t')) \\
&\exp(+i l_{k'k'k'k'} (t-t')) \exp \left( -g^{*}_{k'k'k'k'}(t-t') \right),
\end{split}
\end{equation}
\begin{equation} \label{eq:absorption_modified_Redfield_std}
\begin{split}
&\tilde{A}_{k}(t,t')= \exp(-i E_{k}^0 (t-t')) \\
&\exp(-i l_{kkkk} (t-t')) \exp \left( -g_{kkkk}(t-t') \right),
\end{split}
\end{equation}
and
\begin{equation} \label{eq:Nterm_modified_Redfield_rate_std}
\begin{split}
&N_{k k'}(t,t')=\exp(2i l_{kkk'k'} (t-t')) \\
&\exp ( +2 g_{kkk'k'}(t-t) ) \times \{ \ddot{g}_{k'kkk'}(t-t') \\
&-[\dot{g}_{k'kk'k'}(t-t')-\dot{g}_{k'kkk}(t-t') \\
&+2i l_{k'kk'k'}] \times [\dot{g}_{k'k'kk'}(t-t') \\
&-\dot{g}_{kkkk'}(t-t')+2i l_{k'k'kk'}] \}.
\end{split}
\end{equation}
As in the F\"{o}rster description, the integrand functions of the rate expressions from non-equilibrium and standard treatments
become identical in the limit of large time arguments $t-t'$. 

Furthermore, as in the F\"{o}rster case, the inhomogeneous term vanishes, and the homogeneous component of the population transfer dynamics can be described by introducing 
a population transfer superoperator.
The respective expressions given in Eqs.~(\ref{eq:definition_Theta_Foerster})--(\ref{eq:rate_equation}) are also valid in the Modified Redfield case after replacing $n$ and $m$
with $k$ and $k'$.

\section{Results and discussion} \label{sec:results_discussion}
\subsection{Time-dependence of transfer rates} \label{sec:time_dependence_transfer_rates}

In this section we calculate transfer rates for a model donor--acceptor system motivated by previously studied carotenoid-chlorophyll and carotenoid-purpurin dyads \cite{Perlik2015}. The dyads in Ref. \cite{Perlik2015} are strongly heterogeneous dimers with large donor--acceptor energy gaps compared to the excitonic coupling.
%
%
%
Because we concentrate on studying our newly developed rate theory in this work, we choose the system parameters with a certain freedom to demonstrate the properties of the rates. The parameters of the dyads from Ref. \cite{Perlik2015}, are to be taken as a motivation only. Nevertheless, to distinguish the different characters of the molecules composing the dimer, we refer to them as carotenoid (Car) and chlorophyll (Chl), respectively. All calculations are performed at room temperature, 293 K, and we ignore the so-called static disorder of the transition energies of the molecules. Resonance coupling is set to $J=\unit[-119]{cm^{-1}}$ as in Ref. \cite{Perlik2015}.

Let us first neglect underdamped oscillations characteristic for carotenoid energy gap correlation function. 
Such underdamped oscillations with reorganization energy $\lambda_{UO,i}$, central frequency $\omega_{UO,i}$ and damping constant $\gamma_{UO,i}$ can be included in terms of a spectral density 
\begin{equation} \label{eq:spectral_density_underdamped_oscillator}
J_{UO,i}(\omega)=2 \lambda_{UO,i}  \frac{\omega^2_{UO,i}\omega \gamma_{UO,i}}{(\omega^2-\omega^2_{UO,i})^2+\omega^2\gamma_{UO,i}^2},
\end{equation}
where $i \in \{ \rm{Car},\rm{Chl} \}$. To neglect this spectral density contribution, we set $\lambda_{UO,Car}$ to zero at first, while $\lambda_{UO,Chl}$ is always taken as zero in this work.
We describe the energy gap fluctuations of both components of the dimer by low-frequency overdamped Brownian oscillator spectral densities with reorganization energy $\lambda_{BO,i}$ and damping constant $\Lambda_{BO,i}$ inversely proportional to the decay time $\tau_{BO,i}$
\begin{equation} \label{eq:spectral_density_Brownian_oscillator}
J_{BO,i}(\omega)=2 \lambda_{BO,i} \frac{\omega \Lambda_{BO,i}}{\omega^2+\Lambda_{BO,i}^2}, \; i \in \{ \rm{Car},\rm{Chl} \}.
\end{equation}
We choose $\lambda_{BO,\rm{Car}}=\unit[67]{cm^{-1}}$, $\lambda_{BO,\rm{Chl}}=\unit[60]{cm^{-1}}$, $\tau_{BO,\rm{Car}}=\unit[30]{fs}$ and $\tau_{BO,\rm{Chl}}=\unit[47]{fs}$ for the calculations.
The line shape function components are obtained from the sum spectral density $J_{i}(\omega)=J_{UO,i}(\omega)+J_{BO,i}(\omega)$ via the standard formula
\begin{equation} \label{eq:line_shape_function}
\begin{split}
g_{BO,i}(t)&=\frac{1}{2 \pi} \int^{\infty}_{-\infty} d \omega \frac{1-\cos(\omega t)}{\omega^2} \coth \left( \frac{\omega}{2 k_B T} \right) J_{i}(\omega) 
\\
&+\frac{i}{2 \pi} \int^{\infty}_{-\infty} d \omega \frac{\sin(\omega t)-\omega t}{\omega^2} J_{i}(\omega),
\end{split}
\end{equation}
where $i \in \{ \rm{Car},\rm{Chl} \}$. 
To demonstrate the differences of the time-dependencies of rates calculated under the standard equilibrium and the non-equilibrium bath conditions, 
we first assume the energy gap between the dimer site energies to take the value of $\omega_{21}=\unit[-100]{cm^{-1}}$. This value is comparable with the reorganization energies of the Brownian oscillators.
\begin{figure}[h] 
\includegraphics*[width=\columnwidth]{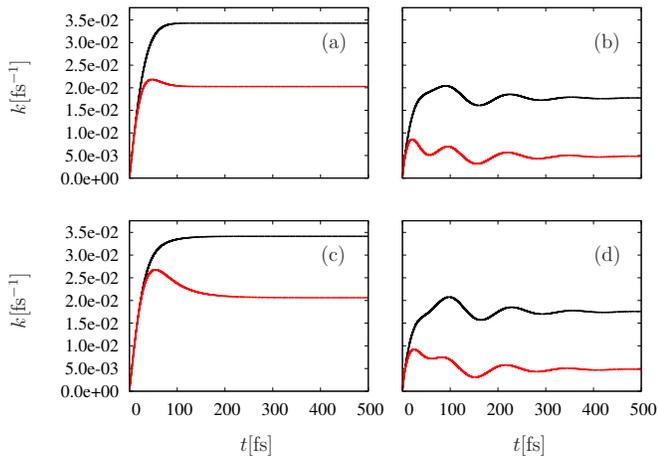}
\caption{\label{fig:combined_figure_time_dependent_rates_without_vibrations}
Upper row: Time-dependence of population transfer rates from standard formulation of (a) F\"{o}rster and (b) Modified Redfield approach.
Lower row: Time-dependence of population transfer rates from non-equilibrium formulation of (c) F\"{o}rster and (d) Modified Redfield approach.
The black and red lines correspond to $k_{1 \to 2}$ and $k_{2 \to 1}$ in the F\"{o}rster case and to $k_{\alpha \to \beta}$ and $k_{\beta \to \alpha}$ in the Modified Redfield case, respectively.
The difference between site energies was chosen as $\omega_{21}=\unit[-100]{cm^{-1}}$.
Intramolecular vibrations of the carotenoid were not taken into account in terms of underdamped oscillators.
The other parameters are specified in Sec.~\ref{sec:time_dependence_transfer_rates}.
}
\end{figure}

Time-dependent rates from both F\"{o}rster and Modified Redfield approach are shown in Fig.~\ref{fig:combined_figure_time_dependent_rates_without_vibrations}. The two theories are formulated in different bases and they refer to transitions between states of different kinds. We denote states localized on individual molecules (local basis) by numbers ($1,2$) and the delocalized eigenstates of the dimer (excitonic basis) by Greek letters ($\alpha,\beta$). For definition of these states see supporting information.
The standard F\"{o}rster theory (stFT) \cite{Yang2002} results in the rates $k_{1 \to 2}=-{\cal R}_{22 11}$ and $k_{2 \to 1}=-{\cal R}_{11 22}$ corresponding to the black and red line
in Fig.~\ref{fig:combined_figure_time_dependent_rates_without_vibrations} (a), respectively. Both rates reach asymptotic values already at about $\unit[100]{fs}$.
These asymptotic values are different from the ones of the rates $k_{\alpha \to \beta}=-{\cal R}_{\beta \beta \alpha \alpha}$ and $k_{\beta \to \alpha}=-{\cal R}_{\alpha \alpha \beta \beta}$ obtained from the standard Modified Redfield theory (stMRT) \cite{Yang2002}
(see Fig.~\ref{fig:combined_figure_time_dependent_rates_without_vibrations} (b)).
Even though the detailed balance condition in the strict sense only applies to the rates obtained from standard Redfield theory (stRT),
the ratio of the asymptotic values of the rates from Fig.~\ref{fig:combined_figure_time_dependent_rates_without_vibrations} (b) exhibits a deviation of less then $\SI{2}{\percent}$ from the thermal population of the exciton states with energy gap of $\unit[260]{cm^{-1}}$ for the given values of $J$ and $\omega_{21}$.
Similar findings are obtained for the time-dependencies of the non-equilibrium F\"{o}rster theory (noneqFT) rates and the non-equilibrium Modified Redfield theory (noneqMRT) rates, as shown in the subfigures (c) and (d) of Fig.~\ref{fig:combined_figure_time_dependent_rates_without_vibrations}, respectively. In the case of noneqMRT 
the deviation from detailed balance is even smaller.
The noneqFT rates take longer to reach their asymptotic values than the ones of the stFT. This effect is clearly recognizable by comparing the back-transfer rates (red lines). Note that the finding of similar asymptotic values for rates from standard and non-equilibrium approach cannot be generalized, as discussed in Sec.~\ref{sec:theoretical_background}.
By comparing Fig. 1b and Fig. 1d one finds differences between noneqMRT and stMRT in the oscillatory dynamics during the early time evolution. Damping of these oscillations indicates bath equilibration. At the upper border of the displayed time interval the rates from noneqMRT and stMRT reach similar values. Close to this upper interval border the oscillations of the rates are almost completely damped out, so that the rates can be approximately considered as asymptotic.
\begin{figure}[h] 
\includegraphics*[width=\columnwidth]{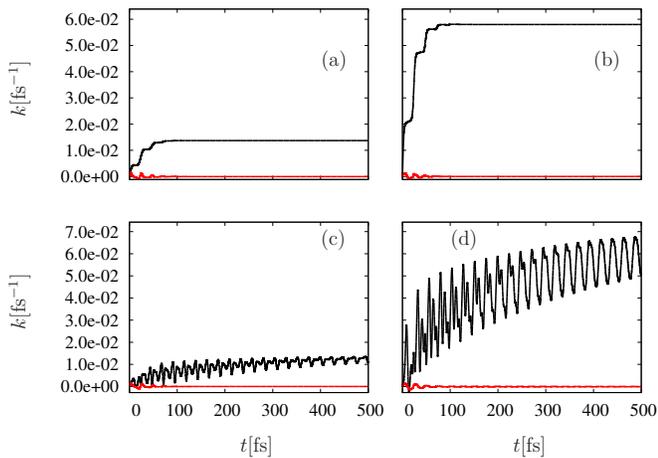}
\caption{\label{fig:combined_figure_time_dependent_rates_with_vibrations}
F\"{o}rster and modified Redfield rates displayed in analogy to Fig.~\ref{fig:combined_figure_time_dependent_rates_without_vibrations},
however with inclusion of intramolecular vibrations in the calculation
(parameters specified in Sec.~\ref{sec:time_dependence_transfer_rates}) 
and under the assumption of an energy gap $\omega_{21}=\unit[-1500]{cm^{-1}}$.
}
\end{figure}

In the next step we include intramolecular vibrational modes characteristic of the carotenoid energy gap correlation function. We introduce an underdamped oscillator with reorganization energy $\lambda_{UO,Car}=\unit[1800]{cm^{-1}}$, vibrational frequency $\omega_{UO,Car}=\unit[1390]{cm^{-1}}$ and damping constant $\gamma_{UO,Car}$ inversely proportional to the decay constant $\tau_{UO,Car}=\unit[200]{fs}$. Our model carotenoid only exhibits one effective high-frequency mode for simplicity. 
We again study the difference between the rates obtained from the standard and our non-equilibrium approaches. We increase the gap between the donor and acceptor site energies to $\omega_{21}=\unit[-1500]{cm^{-1}}$. This energy corresponds more closely to the actual case of the carotenoid-chlorophyll dyad \cite{Perlik2015}. The time-dependencies of the rates are now more influenced by the presence of the intramolecular carotenoid vibrations than by the low-frequency Brownian oscillator contributions of the environment.
The resulting stFT rates are displayed in Fig.~\ref{fig:combined_figure_time_dependent_rates_with_vibrations} (a).
Similar to the results from calculations without vibrations, the respective rates converge toward asymptotic values. Only up to about $\unit[100]{fs}$ rudimentary oscillations are recognizable.
Analogous findings are obtained for the Modified Redfield rates. The rate $k_{\alpha \to \beta}$ reaches a larger asymptotic value than the corresponding F\"{o}rster rate $k_{1 \to 2}$ (see Fig.~\ref{fig:combined_figure_time_dependent_rates_with_vibrations} (b)).
If a non-equilibrium description is chosen, damped oscillations appear throughout the considered time interval (see Fig.~\ref{fig:combined_figure_time_dependent_rates_with_vibrations} (c) and (d)).
These oscillations can be attributed to the intramolecular vibrations of the carotenoid. By considering the evolution of the average of non-equilibrium rates and disregarding the deviations caused by the oscillations, one finds that this average also approaches an asymptotic value. However, convergence of the average rate takes place considerably slower than in the case of standard description, thereby indicating the equilibriation process. 

As the excitonic coupling is much smaller than the reorganization energies of the monomer components, the criterion for applicability of the Modified Redfield approach seems not to be fulfilled at first sight. However, a more careful examination shows that not the size of the reorganization energies by themselves, but rather the size of the related off-diagonal system-bath coupling elements matters for an estimation whether the criterion for applicability of the Modified Redfield approach is fulfilled. The off-diagonal system-bath coupling scales with a product of the coefficients from transformation between localized basis and exciton basis, which become smaller when the energy gap between donor and acceptor increases. Therefore, in the case of the parameter values specified in the discussion above, the requirements for Modified Redfield approximation are likely to apply,  rather than those of F\"{o}rster approximation.

\subsection{Extraction of transfer rates from population dynamics with laser pulse effects} \label{sec:extraction_transfer_rates}

When considering underdamped intramolecular vibrational modes, the transfer rates $k_{2 \to 1}$ and $k_{\beta \to \alpha}$ obtained from the two non-equilibrium approaches may exhibit negative values in the very early time evolution up to about $\unit[20]{fs}$. If this effect is not sufficiently compensated by the complementary rates $k_{1 \to 2}$ and $k_{\alpha \to \beta}$, it can lead to unphysical populations outside the range between $\unit[0]{}$ and $\unit[1]{}$. 
The presence of such unphysical populations seems to be related to the positivity issue, reported previously for Markovian quantum master equations and explained by transient non-Markovian effects before sufficient relaxation of the bath has taken place \cite{Suarez1992,Gaspard1999,Cheng2005}.
These effects can average out under the smoothing influence of additional relaxation channels and excitation by a laser pulse with finite width, as discussed below.
\begin{figure}[h] 
\includegraphics*[width=\columnwidth]{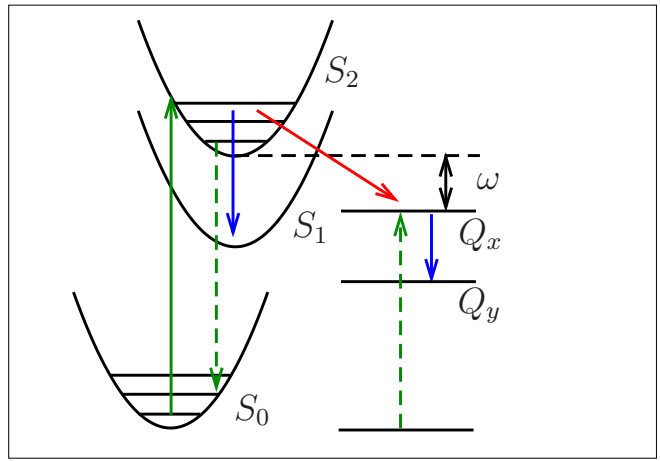}
\caption{\label{fig:level_scheme}
Level scheme of a donor-acceptor complex of carotenoid and chlorophyll, where in F\"{o}rster description after electronic excitation from $S_0$ to $S_2$ intermolecular population transfer from $S_2$ to $Q_x$ is facilitated by resonant emission of the carotenoid and absorption of the chlorophyll component. The fast competing intramolecular population transfer channels $S_2 \to S_1$ and $Q_x \to Q_y$ limit the carotenoid emission to the early stage of the equilibration process in $S_2$.
}
\end{figure}

\begin{figure}[h] 
\includegraphics*[width=\columnwidth]{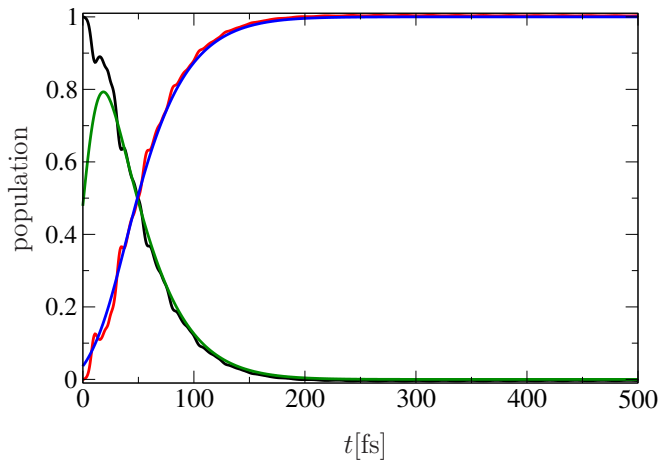}
\caption{\label{fig:demonstration_influence_finite_pulse_width}
Time-dependence of populations $p_{\alpha}$ (black line) and $p_{\beta}$ (red line) resulting from the non-equilibrium Modified Redfield rates 
shown in Fig.~\ref{fig:combined_figure_time_dependent_rates_with_vibrations} together with the corresponding populations extracted from dynamics under the influence 
of a finite excitation pulse (green and blue line).
Note that additional relaxation channels have not been taken into account in this calculation because their influence in decreasing $p_{\alpha}$ and $p_{\beta}$ complicates the comparison.
}
\end{figure}

To describe a donor-acceptor complex of carotenoid and chlorophyll molecules by our dimer model system, we identify the states $\unit[1]{}$ and $\unit[2]{}$ with the $S_2$ state of the carotenoid and the $Q_x$ state of the chlorophyll component, respectively. Decay of the populations in $S_2$ and $Q_x$ accounts for intramolecular population transfer from $S_2$ to $S_1$ and from $Q_x$ to $Q_y$, as sketched in Fig.~\ref{fig:level_scheme}. Those competing population transfer channels allow intermolecular population transfer only at an early stage of the equilibration process in $S_2$.
Furthermore, for the population dynamics, also effects of finite pulse width in the electronic excitation from the carotenoid ground state $S_0$ to $S_2$ play a role.
Under the assumptions that initially only $S_0$ is populated and that the pulses are weak enough not to induce a significant depopulation of $S_0$, the dynamics of the populations $p_{S_2}$, $p_{S_1}$, $p_{Q_x}$ and $p_{Q_y}$ can be expressed in terms of the rates $k_{S_2 Q_x}$, $k_{S_2 S_1}$ and $k_{Q_x Q_y}$ as
\begin{equation}
\begin{split}
  &\dot{\vec{p}}(t')
  =\frac{d}{dt'}
  \left(
    \begin{array}{cccc}
      p_{S_2}(t') \\
      p_{S_1}(t') \\
      p_{Q_x}(t') \\
      p_{Q_y}(t')
    \end{array}
  \right) \\
  &=\left(
    \begin{array}{cccc}
      -k_{S_2 Q_x}(t')-k_{S_2 S_1} & 0 & 0 & 0 \\
      k_{S_2 S_1} & 0 & 0 & 0 \\
      0 & 0 & k_{S_2 Q_x}(t')-k_{Q_x Q_y} & 0 \\
      0 & 0 & 0 & k_{Q_x Q_y}
    \end{array}
  \right) \\
  &\left(
    \begin{array}{cccc}
      p_{S_2}(t') \\
      p_{S_1}(t') \\
      p_{Q_x}(t') \\
      p_{Q_y}(t')
    \end{array}
  \right).
\end{split}
\end{equation}
Inclusion of effects caused by an excitation pulse with time-dependent amplitude $A_{pulse}(t)$ leads to
\begin{equation}
  \vec{p}(t)=\int_{0}^{t} d\tau |A_{pulse}(\tau)|^2 \int_{\tau}^{t} dt' \dot{\vec{p}}(t'); \;\; p_i(\tau)=\delta_{i,S2}.
\end{equation}
To demonstrate the smoothing influence of a finite pulse, the populations $p_{\alpha}$ (black line) and $p_{\beta}$ (red line) 
resulting from the non-equilibrium Modified Redfield rates shown in Fig.~\ref{fig:combined_figure_time_dependent_rates_with_vibrations} are plotted in Fig.~\ref{fig:demonstration_influence_finite_pulse_width} together with the corresponding
populations from dynamics under the influence of a finite excitation pulse (green and blue line). Note that in this calculation, additional relaxation channels have not been taken into account to keep the comparison simple.

If the additional relaxation channels are included, an averaged population transfer rate can be obtained as
\begin{equation} \label{eq:transfer_rate_from_population_dynamics}
  k_{S_2 Q_x,avg}=\lim_{t \to \infty} k_{S_2 S1} \frac{p_{Q_y}(t)}{p_{S_1}(t)}.
\end{equation}
Besides the calculated time-dependent transfer rate $k_{S_2 Q_x}$, additional relaxation channels between $S_2$ and $S_1$ with a time constant of $\unit[95]{fs}$ and between $Q_x$ and $Q_y$ with a time constant of $\unit[20]{fs}$ were assumed. To obtain a realistic description, the FWHM of the pulse was taken as $\unit[19.2]{fs}$, in agreement with the pulse width in the experiment.

\subsection{Dependence of the transfer rates on the donor-acceptor energy gap} \label{sec:frequency_dependence_transfer_rates}

\begin{figure}[h] 
\includegraphics*[width=\columnwidth]{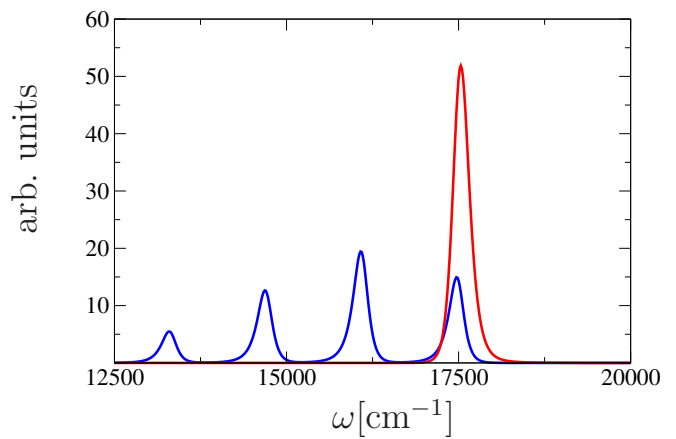}
\caption{\label{fig:em_Car_abs_Chl_lambdaBO_factor_1e-1}
Emission spectrum of carotenoid (blue line) and absorption spectrum of chlorophyll (red line) with parameters specified in Sec.~\ref{sec:time_dependence_transfer_rates}.
}
\end{figure}
Investigation of the relaxation rate dependence on the donor--acceptor energy gap by evaluating Eq.~(\ref{eq:transfer_rate_from_population_dynamics}) allows further interpretation of the underlying processes. 
For the interpretation of the results from the F\"{o}rster approach, it is useful to compare the energy gap dependence of the rates with the overlap integral of the absorption spectrum of the acceptor (chlorophyll) and the emission spectrum of the donor (carotenoid) as a function of the difference between the electronic excitation energies (see Eq.~(\ref{eq:standard_Foerster_rate_product_absorption_emission}) for the respective expression in the time domain). From now on, all energy gap dependencies discussed in this paper will be understood as dependencies on the the energy differences between the localized states of the donor (carotenoid) and acceptor (chlorophyll) molecules. This definition will be followed even in cases where substantial delocalization exists and localized states are not spectroscopically addressable. Due to visual similarity between ordinary absorption spectra and the plots of the energy gap dependence of the spectral overlap and the rates to be presented below, we use the usual spectroscopic vocabulary for their description. For convenience of expression we therefore speak of peaks, bands, their intensities etc. also when describing energy gap dependencies of relaxation rates.

\begin{figure}[h] 
\includegraphics*[width=\columnwidth]{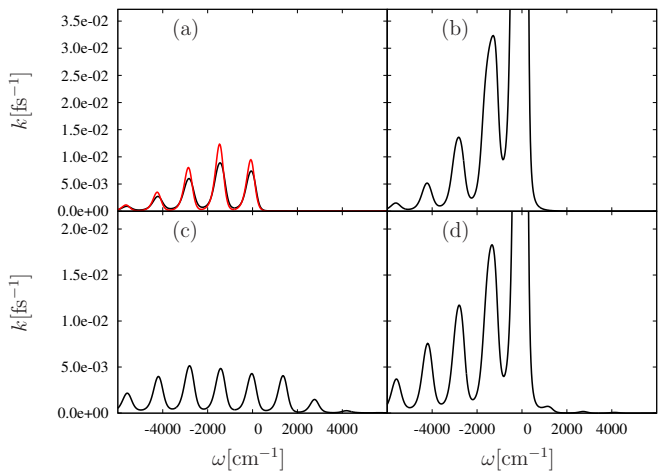}
\caption{\label{fig:combined_figure_frequency_dependence_rates_lambdaBO_factor_1e-1_displayed_frequency_range_changed}
Left column: Energy gap dependence of population transfer rates obtained from standard formulation of (a) F\"{o}rster ($k_{1 \to 2}$) and (b) Modified Redfield approach ($k_{\alpha \to \beta}$).
Together with the F\"{o}rster rate the dependence of the overlap of absorption and emission spectrum from Fig.~\ref{fig:em_Car_abs_Chl_lambdaBO_factor_1e-1} 
as a function of the difference between the electronic excitation energies of acceptor and donor is shown.
Right column: Energy gap dependence of population transfer rates from non-equilibrium formulation of (c) F\"{o}rster and (d) Modified Redfield approach.
The parameters are specified in Sec.~\ref{sec:time_dependence_transfer_rates}.
}
\end{figure}
\begin{figure}[h] 
\includegraphics*[width=\columnwidth]{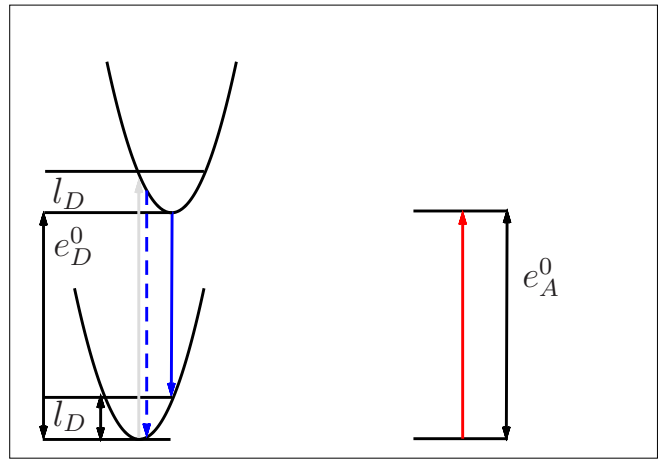}
\caption{\label{fig:sketch_Foerster_transfer}
Potential diagrams of donor (left hand side) and acceptor (right hand side) for the illustration of F\"{o}rster transfer with standard and non-equilibrium treatment.
}
\end{figure}

In Fig.~\ref{fig:em_Car_abs_Chl_lambdaBO_factor_1e-1}, the emission spectrum of our carotenoid model and the absorption spectrum of our chlorophyll model are shown as blue and red lines, respectively. 
The overlap integral as a function of the energy gap between the electronic excitation energies of acceptor and donor is shown in Fig.~\ref{fig:combined_figure_frequency_dependence_rates_lambdaBO_factor_1e-1_displayed_frequency_range_changed} (a) as a red curve. 
In the same subfigure the energy gap dependence of the stFT rate given by Eq.~(\ref{eq:standard_Foerster_rate_homogeneous_term_line_shape_functions}) is displayed as a black curve. 
With respect to the peak positions and relative values of maxima both results agree well.
In the noneqFT rates, the distribution of relative peak intensities changes.
In particular, recognizable peaks also appear at positive values of the difference between the electronic excitation energies of acceptor and donor 
(see Fig.~\ref{fig:combined_figure_frequency_dependence_rates_lambdaBO_factor_1e-1_displayed_frequency_range_changed} (c)).
These findings can be explained in an illustrative way by the potential diagrams shown in Fig.~\ref{fig:sketch_Foerster_transfer}, where donor emission (left hand side) and acceptor absorption (right hand side) are sketched. The solid blue arrow is related to emission in the case of the standard description, whereas the dashed blue arrow illustrates emission from a non-equilibrium initial state.
In the stFT treatment, donor emission is assumed to take place from the equilibrated excited state. Therefore, the largest possible frequency of a vibrational peak in the emission spectrum corresponds to the difference between the minima of the excited state potential and the ground-state potential, i.e.\ to the electronic excitation energy. 
In the acceptor absorption spectrum the energetic position of the single peak corresponds to the electronic excitation energy as well. Thus, an overlap between absorption and emission spectrum can only be obtained if the difference between the electronic excitation energies of acceptor and donor is smaller than zero (or slightly larger than zero by an amount determined by the peak widths).
In the case of noneqFT rates, the emission can take place already at an early stage of bath relaxation, so that the energetic position of vibrational bands in the emission spectrum can become larger than the electronic excitation energy of the carotenoid.
Therefore, peaks of the energy gap dependent rate can also appear at a positive difference between the electronic excitation energies of acceptor and donor.

The dependencies of the Modified Redfield rates on the gap between the electronic excitation energies in the site basis are shown on the right hand side of Fig.~\ref{fig:combined_figure_frequency_dependence_rates_lambdaBO_factor_1e-1_displayed_frequency_range_changed} on the same scale as the corresponding results from the F\"{o}rster approach. 
Even for equal site energies in the localized basis, i.e.\ at an energy gap of zero, the excitonic coupling leads to a splitting between the exciton states, which have delocalized character in this case.
In the limit of energy gaps far exceeding the excitonic coupling the upper and lower exciton state assume the character of the upper and lower localized state, respectively, with only a small mixing contribution from the other localized state. 
Therefore, at energy gaps further to the negative region the results from F\"{o}rster and Modified Redfield description become more similar than in the region close to zero.
In both F\"{o}rster and Modified Redfield rates the relative intensities of the vibrational bands at more negative frequency gaps in the case of non-equilibrium treatment become larger than those from standard treatment. 
This change of the relative intensity of the vibrational bands stems from the dependence of the relative oscillator strengths of transitions from the populated excited state levels to the vibrational levels of the electronic ground state. Which levels are populated depends on whether equilibration has taken place or not.

\begin{figure}[h] 
\includegraphics*[width=\columnwidth]{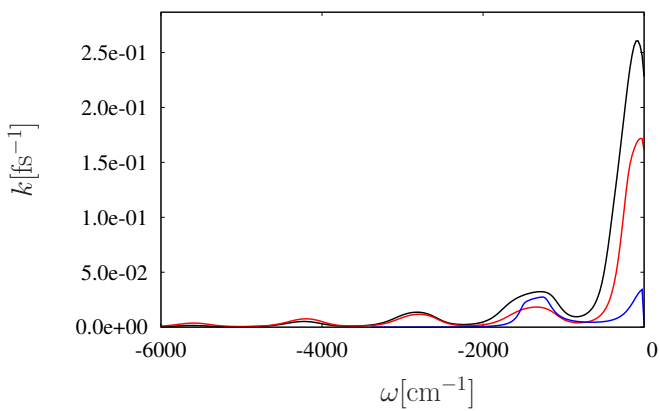}
\caption{\label{fig:combined_figure_frequency_dependence_Redfield_rates_lambdaBO_factor_1e-1}
Energy gap dependence of population transfer rates $k_{1 \to 2}$ obtained from Modified Redfield standard approach (black line), Modified Redfield non-equilibrium approach (red line) and standard Redfield treatment (blue line) with the parameters specified in the caption of Fig.~\ref{fig:combined_figure_frequency_dependence_rates_lambdaBO_factor_1e-1_displayed_frequency_range_changed}.
}
\end{figure}

Besides these similarities, Modified Redfield rates also exhibit remarkable differences compared to the F\"{o}rster rates:
In the region of energy gaps close to zero an intensive band appears, which is cut in the panels on the right hand side of Fig.~\ref{fig:combined_figure_frequency_dependence_rates_lambdaBO_factor_1e-1_displayed_frequency_range_changed} for an easier comparison with the F\"{o}rster rates, but displayed in its full height in the corresponding subfigures of Fig.~\ref{fig:combined_figure_frequency_dependence_Redfield_rates_lambdaBO_factor_1e-1}. 
Such a peak does not appear in the F\"{o}rster rates.
This difference stems from the influence of the factor consisting of line shape function derivative terms in the Modified Redfield rate expressions (see Eqs.~(\ref{eq:modified_Redfield_rate_R}) and (\ref{eq:standard_modified_Redfield_rate_R})). The second-derivative term in this factor corresponds to the correlation function between fluctuations attributed to different exciton states and facilitates transfer between the latter, provided that it contains frequency components resonant with the effective energy gap between the exciton states. This consideration leads to the following conclusion:
Population transfer in the framework of F\"{o}rster theory is facilitated by the resonance coupling between configurations with electronic excitation localized on donor or acceptor, while the bath only plays a role in tuning the energy gap.
In the noneqFT, differences in rates compared to the results from the stFT approach, originate from the inclusion of the bath equilibration process in the excited state of the donor and the resulting modification of the transition frequency of donor emission. In contrast, the bath in the Modified Redfield treatment not only contributes by modifying the effective energy gap between the exciton states during equilibration, but also facilitates population transfer directly by bath DOF explicitly coupling the initial and final states.

In the following we will explain findings in the energy-gap dependence of the Modified Redfield rates by referring to terms in the respective rate equations given in Eqs.~(\ref{eq:modified_Redfield_rate_R}) and (\ref{eq:standard_modified_Redfield_rate_R}). 
In particular, explanation of the intensive band at zero energy gaps in the Modified Redfield rates and of the increased intensities of the neighboring side bands compared to the F\"{o}rster rates requires a closer consideration.
Note, however, that for energy gaps approaching zero the Modified Redfield description becomes less appropriate and that selection of different parameters can lead to significant changes of the relative band intensity in this region.
Nevertheless, the differences between rates from stMRT or noneqMRT description and from stRT description which we are going to discuss in the framework of our model assumptions and parameters are of general validity.
In Fig.~\ref{fig:combined_figure_frequency_dependence_Redfield_rates_lambdaBO_factor_1e-1}, the dependence of stMRT and noneqMRT rates (black and red line, respectively) and of the stRT rate (blue line) on the donor--acceptor energy gap is shown.
While in the stMRT and noneqMRT rates the peak in the region of zero energy gap has an amplitude by about one order of magnitude larger than the side bands, in the stRT rate the amplitude of this band is comparable to the one of the side band at an energy gap corresponding to the underdamped oscillator frequency. 
The broadening of the side band in stRT mainly stems from the extraction of the rate from the population dynamics, whereas in the corresponding bands of the stMRT and noneqMRT rates, a substantial broadening contribution is already included in the rate expressions themselves. 
We will now give some interpretations of features in the energy-gap dependence of the rates by drawing attention to terms in the rate expressions.

In the considered case of zero energy gap between the localized states,
the coefficients for transformation to the exciton basis become equal in absolute value, which indicates maximal delocalization of the exciton states. As a consequence, in the stMRT and noneqMRT rates, the combination of all complex exponentials containing line shape functions, reorganization energies and exciton eigenenergies becomes equal to one. The rate expression integrand reduces to two terms. The first term consists of a second-derivative line shape function expression, i.e.\ a bath correlation function, while the other involves first-derivative line shape function components and reorganization energies. We will denote the first term as ``correlation function term'', and the second term as ``line-shape-function derivative term''. Note that we continue to consider the special case of equal site energies.
The integral over the correlation function term in Eqs.~(\ref{eq:modified_Redfield_rate_R}) and (\ref{eq:standard_modified_Redfield_rate_R}) can be identified with the stRT rate \cite{Pisliakov2006}, as it yields the Fourier component of the correlation function at the excitonic splitting frequency. The line-shape-function derivative term reduces to a time-independent product of reorganization energies in the case of stMRT rates, which is multiplied with a complex exponential oscillating with the excitonic splitting frequency. 
Its integration therefore yields an oscillating contribution to the transfer rate. In the noneqMRT rate the situation is similar, at least under the assumption of a linearization of the imaginary part of line shape functions in high-temperature approximation \cite{Mukamel1995Book}. Note that, in both stMRT and noneqMRT rates, a slight difference between the acceptor and donor site energies can already be sufficient to lead to convergence of the integral expression, because of the additional appearance of real parts of first line-shape-function derivatives which account for the dissipative influence of fluctuations in the transitions between the pure exciton states.
Because the integral over the line-shape-function derivative term does not converge to a constant value in the case of zero energy gap, its interplay with the correlation function term in governing the relaxation dynamics gains importance. In the correlation function term, 
the integration is expected to yield a constant for non-zero temperature at long times. If the respective asymptotic rate contribution is sufficiently large, population transfer takes place fast enough, so that the oscillating rate contribution of the line-shape-function derivative product term contributes significantly only within a finite time window given by the timescale of the resulting population transfer dynamics. Also the influence of competing channels with phenomenological rate constants has a similar effect. At zero energy we thus have two terms of which one is of the same order as the corresponding stRT prediction. Our finding of a relatively large value of the stMRT and noneqMRT rates at zero energy gap can be therefore attributed to the transient influence of the oscillating rate contribution of the line-shape-function derivative term in the course of the relaxation process.

For non-zero energy gap between the localized states the influence of fluctuations on the transfer process enters in first order and leads to peaks at negative-signed energy gaps in the energetic regions of Brownian and underdamped oscillator component. 
Different from the stRT rate, the stMRT and noneqMRT rates exhibit further side bands with progression towards increasingly negative frequency range, as in the F\"{o}rster description. These side bands in the MRT rates can be attributed to the involvement higher-order system-bath coupling terms in the population transfer process. Such effects are taken into account in the framework of the second-order cumulant expansion and enter in terms of the exponentials with line-shape-function arguments in the respective rate expressions.

The vibrational structures in the energy gap dependencies of transfer rates indicate the role of intramolecular vibrations of the carotenoid in establishing resonant energy transfer. While in the case of F\"orster transfer it is obvious that not only the vibrational structure of the excited state, but also the one of the electronic ground state plays a role in this respect, in the Modified Redfield case the situation is more involved. There, the information about coupling between excited state configurations and its connection to the bath dynamics is hidden by the description in the exciton basis. Even though intermediate transitions to the electronic ground state are not commonly associated with the picture of transfer between exciton states, the dynamics of the electronic ground state implicitly enters in the line shape functions. 
For an energy gap of roughly $\unit[-1500]{cm^{-1}}$, which can be considered as a realistic value for the dyad, the noneqMRT approach results in a rate larger than the corresponding F\"{o}rster rate by about a factor of $\unit[2]{}$. 
As mentioned previously, the size of the energy gap relative to the excitonic coupling determines how much the localized states contribute to an exciton state. This participation ratio, which also enters as a scaling of the off-diagonal fluctuations, is quantified by a product of coefficients from transformation between localized basis and exciton basis. In this way it gains influence on the criterion whether the Modified Redfield approach yields an appropriate description. For large enough energy gap, the Modified Redfield approach can even be preferable in cases where the reorganization energy of bath components (including pseudo-modes which enter in terms of underdamped oscillators) are much larger than the excitonic coupling. At the same time, the involvement of bath fluctuations in population transfer for non-zero energy gaps leads to an enhancing influence of such bath components on the transfer rate. Strong system-bath coupling results in an increase of the intensity of side bands in energy gap dependence of the rates (particularly recognizable for lower-order side bands) compared to the F\"{o}rster description. 

Although we do not explicitly account for vibronic coupling, our present study confirms the tendencies reported in  Ref.~\cite{Perlik2015}. In Ref.~\cite{Perlik2015}, larger rates were obtained than estimations based on F\"{o}rster theory predict. As discussed in the present article, such an increased rate can be explained also by Modified Redfield theory. Ground state vibrations were found to play an important role for the transfer efficiency by tuning the energy difference between initial and final state of the transfer process. Such influence is easily understood based on the F\"{o}rster description, however, the same picture is valid in the Modified Redfield description. In the case of resonant donor-acceptor energy gap, involvement of intramolecular vibrational modes enhances energy transfer rates by accepting excess energy corresponding to multiple vibrational quanta. Crucially, we find that our noneqMRT captures this behavior without treating intramolecular vibrational modes explicitly in the Hamiltonian.

\section{Conclusions} \label{sec:Conclusions}

We developed rate description of population transfer under the assumption that the bath degrees of freedom of the donor molecule, after its electronic excitation, are in a non-equilibrium state with respect to the excited state potential energy surface. We derived expressions for the population transfer rates by generalizing the standard F\"{o}rster and Modified Redfield descriptions. 

For a model system similar to a carotenoid-chlorophyll dyad from our previous study, we compared the time-dependencies of energy transfer rates obtained from the standard equilibrium treatment and from our non-equilibrium generalization, and we related the differences with the process of
equilibration. The carotenoid-chlorophyll system is insofar appropriate as an example for application of our derived non-equilibrium description, as the short timescales of intramolecular population transfer through competing channels limit the intermolecular population transfer to an early stage of the bath equilibration. We extracted experimentally relevant rate values from the ratio of asymptotic populations of states populated through competing relaxation channels, and we studied their dependence on the donor-acceptor energy gap. These dependencies reveal and confirm important role of ground state vibrational states of the donor in establishing resonance condition for ultrafast energy transfer. Crucially, these conclusions were reached without the need to explicitly incorporate underdamped vibrational modes into the system Hamiltonian. This was in turn enabled by the non-perturbative character of our non-equilibrium rate description.

\section*{Supplementary material}
See supplementary material for the detailed derivation of the non-equilibrium Modified Redfield rates.

\section*{Acknowledgments}
This work was supported by the Czech Science Foundation (GACR) grant no. 14-25752S and by the Impuls grant in Physics from the Neuron Fund for Support of Science.



\end{document}